\documentclass[structabstract]{aa}  
\newcommand{\kms}{$\mathrm{km\, s^{-1}\, }$}
\usepackage{natbib}
\usepackage{graphicx}
\usepackage[ps2pdf]{hyperref}

\usepackage{txfonts}
\begin{document}

   \title{Kinematic signature of an intermediate-mass black hole in the globular cluster NGC 6388. \thanks{Based on observations collected at the European Organization for Astronomical Research in the Southern Hemisphere, Chile (083.D-0444).}}


   \author{N. L\"utzgendorf
          \inst{1}
          \and
          M. Kissler-Patig\inst{1}
          \and
          E. Noyola\inst{2,3}
          \and
          B. Jalali\inst{1}
		  \and
          P. T. de Zeeuw\inst{1,4}
          \and
          K. Gebhardt\inst{5}          
          \and
          H. Baumgardt\inst{6} 
          }

   \institute{European Southern Observatory (ESO),
              Karl-Schwarzschild-Strasse 2, 85748 Garching, Germany\\
              \email{nluetzge@eso.org}
         \and
             Max-Planck-Institut f\" ur Extraterrestrische Physik, 
             85748 Garching, Germany
         \and
			 University Observatory, Ludwig Maximilians University, 
			 Munich D-81679, Germany	 
         \and
			 Sterrewacht Leiden, Leiden University, 
			 Postbus 9513, 2300 RA Leiden, The Netherlands
         \and
			 Astronomy Department, University of Texas at Austin, 
			 Austin, TX 78712, USA 
         \and
			 School of Mathematics and Physics, University of Queensland, 
			 Brisbane, QLD 4072, Australia}
         	 

   \date{Received Febrary 1, 2011; accepted July 20, 2011}

 
  \abstract
   {Intermediate-mass black holes (IMBHs) are of interest in a wide range of astrophysical fields. In particular, the possibility of finding them at the centers of globular clusters has recently drawn attention. IMBHs became detectable since the quality of observational data sets, particularly those obtained with HST and with high resolution ground based spectrographs, advanced to the point where it is possible to measure velocity dispersions at a spatial resolution comparable to the size of the gravitational sphere of influence for plausible IMBH masses.}
   {We present results from ground based VLT/FLAMES spectroscopy in combination with HST data for the globular cluster NGC 6388. The aim of this work is to probe whether this massive cluster hosts an intermediate-mass black hole at its center and to compare the results with the expected value predicted by the $M_{\bullet} - \sigma$ scaling relation.}
   {The spectroscopic data, containing integral field unit measurements, provide kinematic signatures in the center of the cluster while the photometric data give information of the stellar density. Together, these data sets are compared to dynamical models and present evidence of an additional compact dark mass at the center: a black hole.}
   {Using analytical Jeans models in combination with various Monte Carlo simulations to estimate the errors, we derive (with $68 \% $ confidence limits) a best fit black-hole mass of $ (17 \pm 9) \times 10^3 M_{\odot}$ and a global mass-to-light ratio of $M/L_V = (1.6 \pm 0.3) \ M_{\odot}/L_{\odot}$.}
   {}

   \maketitle
%

\section{Introduction}


For a long time, only two mass ranges of black holes were known. On the one hand, we have stellar mass black holes, which are remnants of massive stars, and can be observed in binary systems. On the other hand, there are supermassive black holes at the centers of galaxies, some of them accreting at their Eddington limit and producing the brightest objects known (quasars).  It has been demonstrated that supermassive black holes show a tight correlation between their mass and the velocity dispersion of the galaxy in which they reside \citep[e.g.][]{ferrarese_2000,gebhardt_2000, gultekin_2009}. Extrapolating this relation to the lower velocity dispersions of globular clusters, with $\sigma \sim 10 - 20$ \kms, predicts central black holes in these objects with masses of $10^3 - 10^4 M_{\odot}$. 

Due to the small amount of gas and dust in globular clusters, the accretion efficiency of a potential black hole at the center is expected to be low. Therefore, the detection of IMBHs at the centers of globular clusters through X-ray and radio emissions is challenging \citep{miller_2002, maccarone_2008}. Nevertheless, there is another way to detect IMBHs in globular clusters: exploring the kinematics of these systems in the central regions. This method, proposed forty years ago \citep{bahcall_1976,wyller_1970}, has long been limited by the quality of observational datasets, since it requires velocity dispersion measurements at a spatial resolution comparable to the size of the gravitational sphere of influence for plausible IMBH masses ($1 - 2 ''$ for large Galactic globular clusters). However, with existence of the Hubble Space Telescope (HST) and with high spatial resolution ground based integral-field spectrographs, the search for IMBHs was revitalized.

\cite{gebhardt_1997,gebhardt_2000} and \cite{gerssen_2002} claimed the detection of a black hole of $(3.2 \pm 2.2) \times 10^3 M_{\odot}$ in the globular cluster M15 from photometric and kinematic observations. After more investigations this cluster no longer appears as a strong IMBH candidate \citep[e.g.][]{dull_1997,baumgardt_2003a,baumgardt_2005,bosch_2006}, but new detections of IMBH candidates in other clusters followed. \cite{gebhardt_2002,gebhardt_2005} used the velocity dispersion measured from integrated light near the center of the M31 cluster G1 to argue for the presence of a $(1.8 \pm 0.5) \times 10^4 M_{\odot}$ dark mass at the cluster center. The possible presence of an IMBH in G1 gained further credence with the detection of weak X-ray and radio emission from the cluster center \citep{pooly_2006, kong_2007,ulvestad_2007}. Also, the globular cluster $\omega$ Centauri (NGC 5139) has been proposed to host a black hole at its center \citep{noyola_2008, noyola_2010}. The authors measured the velocity-dispersion profile with an integral field unit and used orbit based dynamical models to analyze the data. \cite{vdm_2010} studied the same object using proper motions from HST images. They found less compelling evidence for a central black hole, but more importantly, they found a location for the center that differs from previous measurements. Both G1 and $\omega$ Centauri have been suggested to be stripped nuclei of dwarf galaxies \citep{freeman_1993, meylan_2001} and therefore may not be the best representatives of globular clusters. The key motivation of this work is to probe more globular clusters for the presence of IMBHs.

Further evidence for the existence of IMBHs is the discovery of ultra luminous X-ray sources at non-nuclear locations in starburst galaxies \citep[e.g.][]{fabbiano_1989, colbert_1999, matsumoto_2001, fabiano_2001}. The brightest of these compact objects (with $L \sim 10^{41} \ \mathrm{erg}\, \mathrm{s}^{-1}$) imply masses larger than $10^3 M_{\odot}$ assuming accretion at the Eddington limit. Several realistic formation scenarios of black holes in globular clusters have been developed. The two main formation theories are: a) IMBHs would be Population III stellar remnants \citep{madau_2001}, or b) they would form in a runaway merging of young stars in sufficiently dense clusters \citep{zwart_2004, gurkan_2004, freitag_2006}. In addition \cite{miller_2002} presented scenarios for the capture of clusters by their host galaxies and accretion in the galactic disk in order to explain the observed bright X-ray sources.

Our goal in this paper is to study the globular cluster NGC 6388. This cluster is located $11.6$ kpc away from the Sun, in the outer bulge of our Galaxy. \cite{white_1972} assigned it a high metallicity after studying its color magnitude diagram. Later, \cite{illingworth_1974} determined the dynamical mass of the cluster: with $\sim 1.3 \times 10^6 M_{\odot}$ it belongs to the most massive clusters in the Milky Way. In addition, its high central velocity dispersion of $18.9$ \kms  \citep{pryor_1993}, and assuming a black-hole mass correlation with velocity dispersion, make NGC 6388 a good candidate for detecting an intermediate-mass black hole. Besides the kinematic properties, the photometric characteristics are also quite interesting. \citet[hereafter NG06]{noyola_2006} found a shallow cusp in the central region of the surface brightness profile of NGC 6388. N-body simulations showed that this is expected for a cluster hosting an intermediate-mass black hole \citep{baumgardt_2005}. For that reason, \cite{lanzoni_2007} investigated the projected density profile and the central surface brightness profile with a combination of HST high-resolution and ground-based wide-field observations. They found the observed profiles are well reproduced by a multimass, isotropic, spherical King model, with an added central black hole with a mass of $\sim 5.7 \times 10^3 M_{\odot}$. Also, the work of \cite{miocchi_2007} suggests the presence of an intermediate-mass black hole in NGC 6388 as a possible explanation for the extended horizontal branch. 

Another interesting fact about NGC 6388 is that it appears to contain multiple stellar populations \cite[e.g.][]{yoon_2000,piotto_2008}. This fact makes the scenario of merging clusters plausible, leading to potentially complicated dynamics in the center. Recently, X-ray observations for NGC 6388 showed no significant signatures for an accreting IMBH  \citep{cseh_2010}. But as already mentioned, this does not rule out a quiescent black hole. In summary, this cluster was chosen as it presents many interesting features. We measured the kinematics of the central regions, allowing us to probe the result of \cite{lanzoni_2007} with a different method, taking the kinematic properties as well as the photometric properties into account.

The basic approach of this work is to first study the light distribution of the cluster. Photometric analysis, such as the determination of the cluster center and the measurement of a surface brightness profile, is described in section \ref{phot}. De-projecting this profile gives an estimation of the gravitational potential produced by the visible mass. The next step is to study the dynamics of the cluster. Section \ref{spec} gives an overlook of the FLAMES observations and data reduction and section \ref{kin} describes the analysis of the spectroscopic data. With the resulting velocity-dispersion profile, it is possible to estimate the actual dynamical mass. The next step is to compare the data to dynamical Jeans models (section \ref{jeans}). These models take the light profile and predict a velocity-dispersion profile, which is scaled to the data in order to obtain the mass-to-light ratio. This is done for models containing different black-hole masses until the best fit to the observed profile is found. In section \ref{con} we summarize our results, list our conclusions and give an outlook for further studies. 

\begin{table}
\caption{Properties of the globular cluster NGC 6388 from the references: NG=\cite{noyola_2006}, H= \cite{harris_1996}, M= \cite{moretti_2009}, L=\cite{lanzoni_2007} and PM=\cite{pryor_1993}.}             
\label{tab_6388}      
\centering
\begin{tabular}{l l l}
\hline \hline
\noalign{\smallskip}
 Parameter & Value & Reference\\
 \noalign{\smallskip}
\hline
\noalign{\smallskip}
 RA (J200) & $17\mathrm{h} \ 36\mathrm{m} \ 17\mathrm{s}$  & NG \\ 
 DEC (J200) & $-44^{\circ}\ 44' \ 08''$ & NG\\ 
Galactic Longitude l & $345.56 °$ &H\\
Galactic Latitude b & $-6.74 °$ &H\\
Distance from the Sun $R_{\mathrm{SUN}}$ & $11.6 \ \mathrm{kpc}$ & M \\
Core Radius $r_c$ & $7.2''$ & L  \\
Central Concentration c & $1.8$ & L  \\
Heliocentric Radial Velocity V$_r$ & $81.2 \pm 1.2 \ \mathrm{km}/\mathrm{s}$ & H \\ 
Central Velocity Dispersion $\sigma$ & $18.9 \ \mathrm{km}/\mathrm{s}$ & PM \\
Age & $ (11.5 \pm 1.5) \ \mathrm{Gyr}$ & M \\
Metallicity $[\mathrm{Fe}/\mathrm{H}]$ & $-0.6 \ \mathrm{dex}$ & H \\
Integrated Spectral Type  & G2 &H \\ 
Reddening E(B-V) & $0.38 $ & M \\
Absolute Visual Magnitude $M_{Vt}$ & $-9.42 \ \mathrm{mag}$ & H  \\
\noalign{\smallskip}
\hline 
\end{tabular} 
\end{table}


\section{Photometry} \label{phot}


The photometric data, retrieved from the archive, were obtained with the Advanced Camera of Surveys (ACS) of the Hubble Space Telescope (HST) in the Wide-Field Channel (under the HST program SNAP-9821, PI: B.J. Pritzl) between October 2003 and June 2004. This data set is composed of six B (F435W), V (F555W) and I (F814W) images with exposure times of 11, 7 and 3s, respectively. It gives a complete coverage of the central region of the cluster out to a radius of $110''$. The data were calibrated, geometrically corrected and dither-combined as retrieved from the European HST-Archive (ST-ECF, Space Telescope European Coordinating Facility\footnote{Based on observations made with the NASA/ESA Hubble Space Telescope, and obtained from the Hubble Legacy Archive, which is a collaboration between the Space Telescope Science Institute (STScI/NASA), the Space Telescope European Coordinating Facility (ST-ECF/ESA) and the Canadian Astronomy Data Centre (CADC/NRC/CSA).}).

\subsection{Color magnitude diagram (CMD) of NGC 6388} \label{sec_cmd}

The CMD was obtained using the programs \textit{daophot II}, \textit{allstars} and \textit{allframes} by P. Stetson, applied to our HST image. For a detailed documentation of these routines see \cite{stetson_1987}. These programs were especially developed for photometry in crowded fields and therefore ideally suited for the analysis of globular clusters. The routines \textit{find}, \textit{phot} and \textit{psf} identify the stars, perform aperture photometry and compute the average point spread function (PSF) over the image, respectively.

Once the PSF has been defined, the next step is to group the neighboring stars to apply the multiple-profile-fitting routine simultaneously by the task \textit{allstars}. Afterwards, the \textit{find} task is applied again to find, in the star-subtracted image, stars which were not found in the first run. The entire procedure was performed on the V- and I-band images independently. As a next step the programs \textit{daomaster} and \textit{allframes} were used to combine both images and to create the final catalog containing all stars, their positions and magnitudes in the two bands. At the end, we calibrated the final catalog to the Johnson magnitude system by following the steps described in \cite{sirianni_2005}.

\begin{figure}
   \centering
   \includegraphics[width=0.5\textwidth]{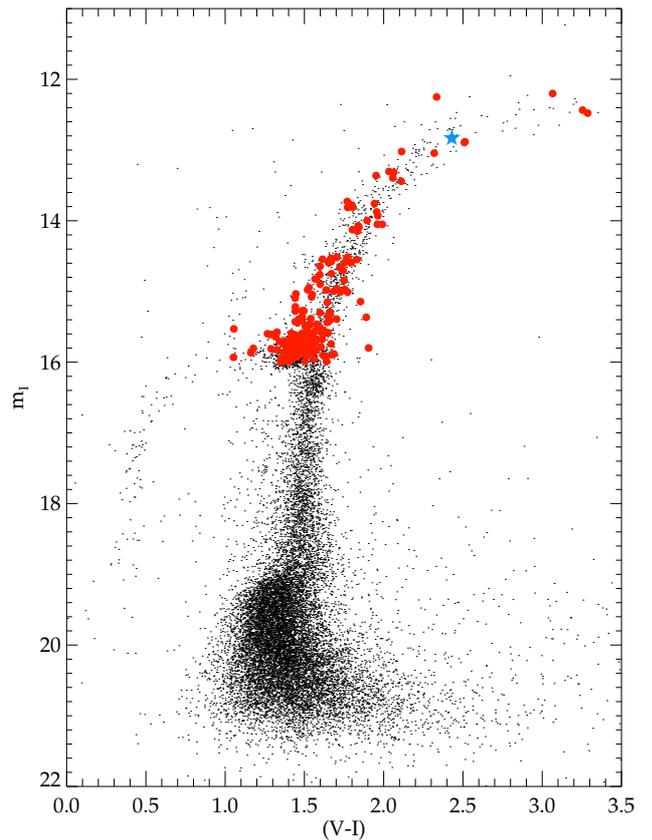}
      \caption{The color-magnitude diagram of NGC 6388. Overplotted are the brightest stars identified in the ARGUS field of view (red circles), and the template star used (star symbol).}
         \label{cmd}
   \end{figure}

In order to get better quality at the faint end of the CMD, one final step was applied to the catalog. The program \textit{separation} \citep{stetson_2003} computes a separation index for every star in a catalog. This index is calculated by, first, evaluating a local surface brightness at the position of a given star. Second, the local surface brightness is compared to the sum of the surface brightnesses produced by the PSF of all the other stars in the field at the position of the centroid of that star. The ratio of these two surface brightness values expressed in magnitudes determines the separation index. Thus, the stars could be selected considering background-light contamination and not only by magnitude. Figure \ref{cmd} shows the final CMD of stars with a separation index $\ge 5$ overplotted with the positions of the brightest stars in the ARGUS pointing and the spectroscopic template star (see section \ref{spec}). 

\subsection{Center of the cluster} \label{phot_center}

Using the star catalog generated with \textit{daophot}, the center of the cluster can be determined. Precise knowledge of the cluster center is important since the shape of the surface brightness and the angular averaged line-of-sight velocity distribution (LOSVD) profiles depend on the position of that center. Using the wrong center typically produces a shallower inner profile. For example determining the center for $\omega$ Centauri is not trivial \citep{noyola_2006,noyola_2010,vdm_2010}. Fortunately, NGC 6388 is not as extended and has a steeper light profile than $\omega$ Centauri so that the center is easier to determine. NG06 determined the center of this cluster to be at $\alpha=17:36:17.18, \ \delta=-44:44:07.83 \ (J2000)$, with an uncertainty of $0.5''$, by minimizing the standard deviation of star counts in eight segments of a circle. In view of the discrepancies about the center location for $\omega$ Centauri, we decided to recompute the center of NGC 6388 and to evaluate how precisely the center can be determined. NG06's center on our I-band image was used as a first guess for the following methods to determine the center in the ACS images, i.e. our reference frame for the spectroscopy. 

   \begin{figure*}
   \centering
   \includegraphics[width=\textwidth]{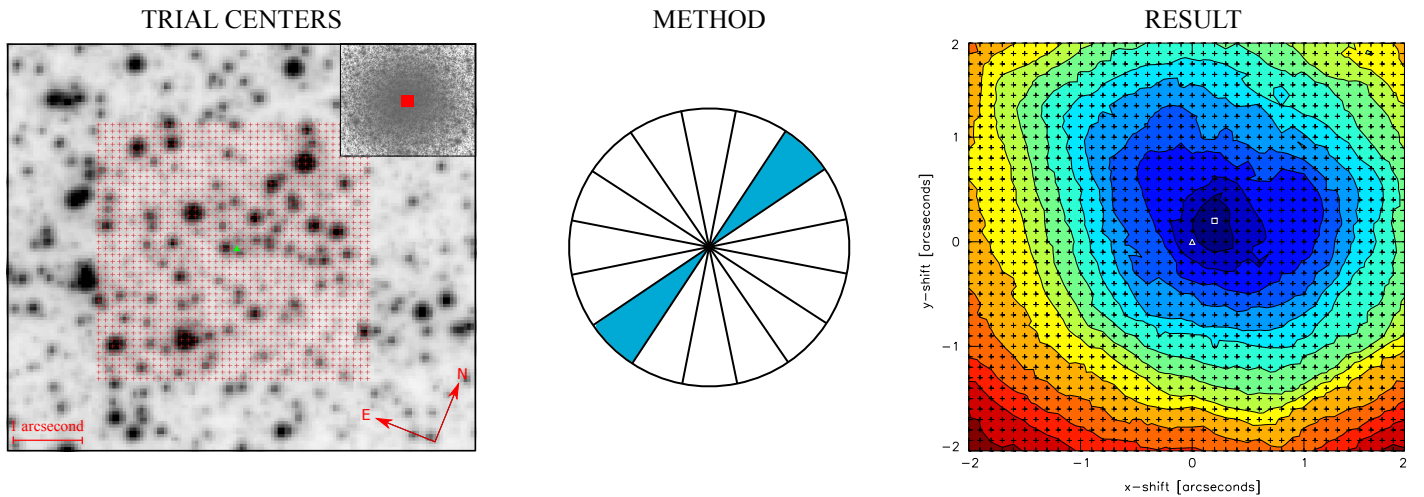}
      \caption{The method to determine the globular cluster center. From left to right: a zoom into our HST/ACS image of NGC 6388 (used to derive the CMD and luminosity profile for the cluster) with the over-layed grid of trial centers. Around each trial center, differential stellar counts and stellar luminosity were computed in two opposite wedges extending 15'' in radius [middle panel]. A minimization of the residuals, as shown in the contour plot [right panel] determines the center to within $0.2''$. The distances between the contours vary between 40 - 70 stars for the star count methods.}
         \label{center}
   \end{figure*}

The first method is a simple star count as described in \cite{47tuc}. The catalog generated with \textit{daophot} (see section \ref{sec_cmd}) contains 88,406 stars.  In a field of $4'' \times 4''$ a grid of trial centers was created, using a grid spacing of 2 ACS pixels ($0.1''$). Around each trial center a circle of 300 pixels ($15''$) was considered and divided into 16 wedges as shown in Figure \ref{center}. The stars in each wedge were counted and compared to the opposite wedge. The differences in the total number of stars between two opposing wedges were summed for all 8 wedge pairs. The coordinates that minimized the difference defined a first guess of the center of the cluster. This center was refined as described in the next sections. We compared our result to the center obtained by NG06. The two centers are only $0.32''$ apart and thus coincide within the error bars of $0.5''$ (as determined by NG06 performing artificial image tests). 

The second method that we used is also described in \cite{47tuc}. Instead of comparing the total number in pairs of opposite wedges, a cumulative radial distribution for each wedge in 4 bins was generated. This time 8 wedges were used instead of 16 to avoid too large stochastic errors due to insufficient  numbers of stars in each bin. The bins were placed at equidistant radii. Again the absolute value of the integrated difference between the radial distributions in any two opposing wedges was calculated and the minimum used to determine the new center of the cluster. The so derived center lies within $0.1''$ of the one derived with the previous method and within $0.3''$ of NG06's location.

We present a last method in which the light of the stars instead of their number, is considered. Similar to the first method, 16 wedges without any radial bins were generated, but this time not the stars were counted but the luminosities of the stars were summed up in each wedge. In order to avoid a bias by the contamination of a few bright stars, only stars from the lower giant branch and the horizontal branch (between $m_V = 15$ and $m_V = 19$) were used. This approach reproduces NG06's center within the error bars ($0.28''$) as well. The method is illustrated in Figure \ref{center}. Shown are the grid of trial centers, the 16 wedges we applied to each of them, and the final contour plot.

For the subsequent analysis, we used the center derived from our ACS catalog. To estimate the error of our center determination we took the scatter from the three different methods as well as an additional test where only eight wedges where used. For these eight wedges we repeated the routine by only using the cardinal wedges and the semi-cardinal wedges separately. For this dataset (HST J8ON08OYQ, also used as position reference coordinates) we derived a final position of the center of:

\begin{equation} (x_c,y_c) = (2075.5, 2548.5) \pm (3.1,2.1) \ \mathrm{pixel} \end{equation} \label{p3}
\begin{eqnarray} \alpha &=& 17:36:17.441,  \ \Delta\alpha = 0.2''  \ \mathrm{(J2000)} \\ \label{p4}
				 \delta &=& -44:43:57.33,  \ \Delta\delta = 0.1'' \end{eqnarray} \label{p4}

All the derived centers lie within a few tenths of an arcsecond ($\sim 10^{-2}$ pc) radius. However, it must be considered that all the methods are using the same catalog derived by DAOPHOT. This catalog most likely suffers from incompleteness since the bright stars are covering the fainter stars underneath. This could bias the center towards the bright stars even if they are excluded in the counts. Our center agrees within the error bars with the one determined in NG06 on WFPC2 images. It also coincides with the center found by \cite{lanzoni_2007} which, according to them, lies $\sim 0.5''$ northwest of the NG06 center and therefor closer to ours.

\subsection{Surface brightness profile} \label{phot_sb}

The last step of the photometric analysis was to obtain the surface brightness profile. This is required as an input for the Jeans models described in the following section. To obtain the profile, a simple method of star counts in combination with an integrated light measurement from the ACS image was applied. The fluxes of all stars brighter than $m_V = 18$ were summed in radial bins of equal width (50 pix) around the center and divided by the area of each bin. Since there are no stars in the gap between the two ACS CCD chips this area was subtracted from each affected bin.  

In addition, the integrated light for stars fainter than $m_V = 18$ was measured directly from the HST image. Using the same radial bins as in the star count method, we measured the statistical distribution of counts per pixel excluding regions with stars with $m_V < 18$. After trying different methods to derive the ``average" of the distribution of counts, we applied a simple average that takes into account the faintest pixels i.e. stars. At the end, the flux per pixel was transformed back into magnitude per square arcseconds and added to the star counts profile.

We compared our profile with \cite{trager_1995}, \cite{lanzoni_2007} and NG06. The latter was derived by measuring the integrated light using a bi-weight estimator while \cite{lanzoni_2007} derived their points by taking the average of the counts per pixel in each bin. We were able to reproduce the shape in the outer regions, but due to the method and data that we used our profile in the innermost region has a high uncertainty. The errors on our profile were obtained by Poisson statistics of the number of stars in each bin. With a linear fit inside the core radius ($\sim 7''$) we derived a logarithmic slope of $\beta = 0.7 \pm 0.2$ (where $\beta$ corresponds to $\mu_V \sim \log r^{\, \beta}$ with the surface brightness $\mu_V$), which results in a slope of the surface luminosity density $I(r) \varpropto r^{\, \alpha}$ of $\alpha = -0.28 \pm 0.08$. This value is steeper (but consistent within the errors) than the slope of $\alpha =-0.13 \pm 0.07$ derived by NG06 and also consistent with the slope derived by \cite{lanzoni_2007} of $\alpha=-0.23 \pm 0.02$.

\begin{figure}
  \centering
   \includegraphics[width=9 cm]{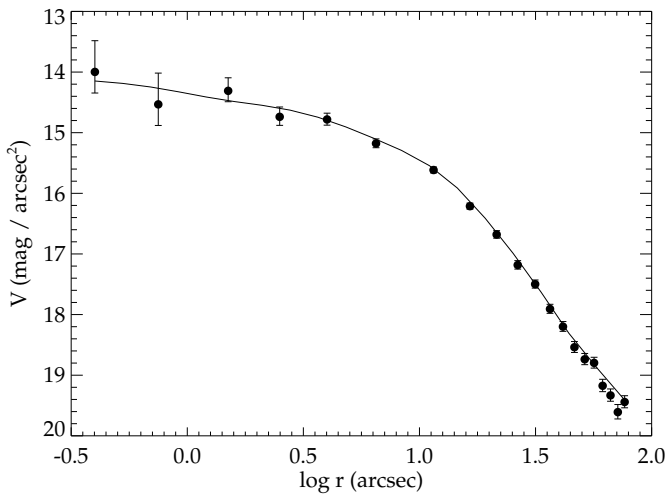}
      \caption{The surface brightness profile of NGC 6388. The profile shows a clear cusp for the inner 10 arcseconds. Also shown is the MGE fit (solid line) which was used to parametrize our profile (see section \ref{models}.)}
         \label{sb_fit}
   \end{figure}

To derive the SB profile, we used a calibrated catalog but we also combined this with measurements that we obtained from counts per pixel. It is possible that the transformations from counts into magnitudes do not match the calibrations of the catalog and therefore would cause an offset in our profile. The fact that we used all the light in the image, including the background light could also shift the profile to higher magnitudes than the intrinsic brightness. For that reason, we scaled our profile to the outer parts of the profile by \cite{trager_1995}. This profile was obtained with photometrically calibrated data and provides a good reference for the absolute values of the profile. The final result of the surface brightness profile is shown in Figure \ref{sb_fit}.

\begin{table}
\caption{The derived surface brightness profile in the V-band. $\Delta V_{h}$ and $\Delta V_{l}$ are the high and low values of the errors, respectively.}             
\label{tab_sb}      
\centering
\begin{tabular}{c c c c}
\hline \hline
\noalign{\smallskip}
  
$\log r$ 	& $V$  								&$\Delta V_{l}$ 					& $\Delta V_{h}$ \\ 

[arcsec] 			& $[\mbox{mag} / \mbox{arcsec}^2]$ 	&$ [\mbox{mag} / \mbox{arcsec}^2]$ 	& $[\mbox{mag} / \mbox{arcsec}^2]$ \\ 
\noalign{\smallskip}
\hline
\noalign{\smallskip}

    - 0.40   &   14.00  &   0.52  &  0.35  \\
    - 0.12   &   14.53  &   0.52  &  0.35  \\
      0.18   &   14.31  &   0.21  &  0.18  \\
      0.40   &   14.74  &   0.16  &  0.14  \\
      0.60   &   14.78  &   0.10  &  0.09  \\
      0.81   &   15.18  &   0.07  &  0.07  \\
      1.06   &   15.62  &   0.05  &  0.05  \\
      1.22   &   16.21  &   0.05  &  0.05  \\
      1.33   &   16.68  &   0.06  &  0.06  \\
      1.42   &   17.18  &   0.07  &  0.07  \\
      1.50   &   17.50  &   0.07  &  0.07  \\
      1.56   &   17.91  &   0.08  &  0.07  \\
      1.62   &   18.20  &   0.08  &  0.08  \\
      1.67   &   18.54  &   0.09  &  0.09  \\
      1.71   &   18.74  &   0.10  &  0.09  \\
      1.75   &   18.80  &   0.09  &  0.09  \\
      1.79   &   19.17  &   0.11  &  0.10  \\
      1.82   &   19.33  &   0.11  &  0.10  \\
      1.85   &   19.61  &   0.13  &  0.11  \\
      1.88   &   19.44  &   0.10  &  0.10  \\
      
\noalign{\smallskip}
\hline
\end{tabular}
\end{table}

\section{Spectroscopy} \label{spec}

\subsection{Observations}

\begin{figure}
   \centering
   \includegraphics[width=0.45\textwidth]{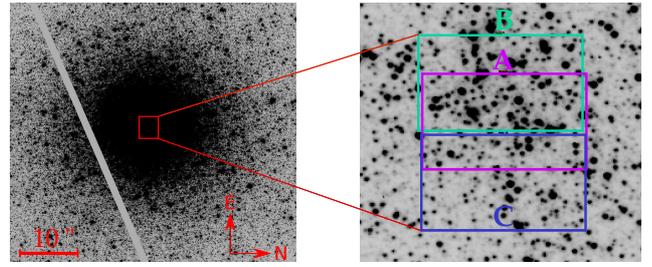}
      \caption{The positions of the three ARGUS pointings (A, B and C) reconstructed on the HST/ACS image.}
         \label{recon}
   \end{figure}

The spectroscopic data were observed with the GIRAFFE spectrograph of the FLAMES (Fiber Large Array Multi Element Spectrograph) instrument at the Very Large Telescope (VLT) in ARGUS (Large Integral Field Unit) and IFU (Integral Field Unit) mode. The set contains spectra from the center (ARGUS) and the outer regions (IFU) for the globular cluster NGC 6388. The observations were performed during two nights (2009-06-14/15). The ARGUS unit was set to the 1 : 1 magnification scale (pixel size: $0.3 ''$, $14 \times 22$ pixel array) and pointed to  three different positions (exposure times of the pointings: A:  $3 \times 480 \, \mathrm{s} + 3 \times 1500 \, \mathrm{s}$, B: $3 \times 1500 \, \mathrm{s}$, C: $1 \times 1500 \, \mathrm{s}$) to cover the sphere of influence of the potential black hole (see Figure \ref{recon}). The IFU fibres were placed around the core radius of the cluster to obtain velocity dispersions in the outer regions.

The kinematics were obtained from the analysis of the Calcium Triplet ($\sim 850 \, \mathrm{nm}$) which is a strong feature in the spectra. The expected velocity dispersions lie in the range 5-20 \kms and need to be measured with an accuracy of 1-2 \kms. This implied using a spectral resolution around $10000$, available in the low spectral resolution mode set-up LR8 ($820-940 \, \mathrm{nm}, \, \mathrm{R} = 10400$).

\subsection{Data reduction}

We reduced the spectroscopic data with the GIRAFFE pipeline programmed by the European Southern Observatory (ESO). This pipeline consists of five recipes which are briefly described below. 

First, a master bias frame was created by the recipe \textit{gimasterbias} from a set of raw bias frames. Next, a master dark was produced by the recipe \textit{gimasterdark} which corrected each input dark frame for the bias and scaled it to an exposure time of 1 second. The recipe \textit{gimasterflat} was responsible for the detection of the spectra on a fiber flat-field image for a given fiber setup. \textit{Gimasterflat} located the fibers, determined the parameters of the fiber profile by fitting an analytical model of this profile to the flat-field data and created an extracted flat-field image. This image was later used to apply corrections for pixel-to-pixel variations and fiber-to-fiber transmission.

The pipeline recipe \textit{giwavecalibration} computed a dispersion solution and a slit geometry table for the fiber setup in use. This was done by extracting the spectra from the bias corrected ThAr lamp frame, using the fiber localization (obtained through the flat-field), selecting the calibration lines from the line catalog, and predicting the positions of the ThAr lines on the detector using an existing dispersion solution as an initial guess. The quality of the calibration was checked in two ways. The first one was by measuring the centroid of one dominating sky emission line in all 14 sky spectra and the second was a cross-correlation in Fourier-space of all sky spectra. 

Both methods did not show any systematic shift of the spectra and resulted in an RMS of $0.03$ \AA, which corresponds to a velocity of $1$ \kms. The last recipe, \textit{giscience} combined all calibrations and extracted the final science spectra. A simple sum along the slit was applied as extraction method. The input observations were averaged and created a reduced science frame, the extracted and rebinned spectra frame. At the end, the recipe also produced a reconstructed image of the respective field of view of the IFU and the ARGUS observations.

The most important parts of our reduction are the sky subtraction and an accurate wavelength calibration to avoid artificial line broadenings due to incorrect line subtractions. The program we used was developed by Mike Irwin and described in \cite{battaglia_2008}. To subtract the sky, the program first combined all (14) sky fibers using a 3-sigma clipping algorithm and computed an average sky spectrum. Using a combination of median and boxcar, it then split the continuum and the sky-line components in order to create a sky-line template mask. For the object spectra, the same method of splitting the continuum from the lines was applied. In the process, the sky-lines were masked out to get a more accurate definition of the continuum. Afterwards, the sky-line mask and the line-only object spectra were compared to find the optimum scale factor for the sky spectrum in order to subtract the sky-lines from the object spectra. The continuum was added back to the object spectra and the sky continuum subtracted by the same scaling factor as obtained for the lines (assuming that lines and continuum have the same scaling factor).

After applying the sky subtraction program, the last step in the reduction of the data was to remove the cosmic rays from the spectra. This was done using the program LA-Cosmic developed by \cite{Lacos} and based on a Laplacian edge detection. In order to avoid bright stars dominating the averaged spectra when they were combined, we applied a simple normalization by fitting a spline to the continuum and dividing by it. 

For the small IFUs, we applied the same reduction steps as for the large integral-field unit, except for the cosmic ray removal, since LA-Cosmic did not give a satisfying result. However, we were able to average all exposures applying a sigma clipping method in order to remove the cosmic rays, due to the fact that these pointings were not dithered.

     \begin{figure}
   \centering
   \includegraphics[width=0.5\textwidth]{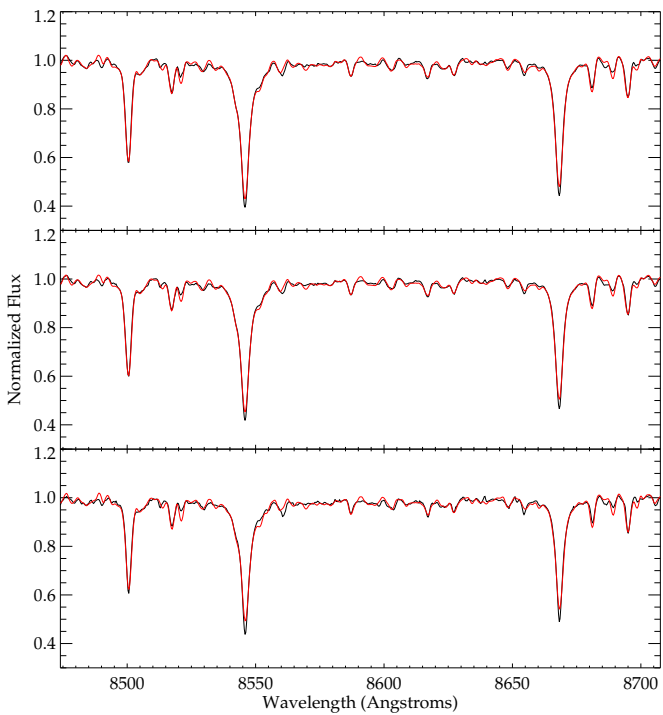}
      \caption{Combined spectra of the first, third and sixth bin overplotted by their best fit (red line). Due to the binning, the spectra show a high signal-to-noise ratio.}
         \label{model_point}
   \end{figure}

      \begin{figure*}
   \centering
   \includegraphics[width=\textwidth]{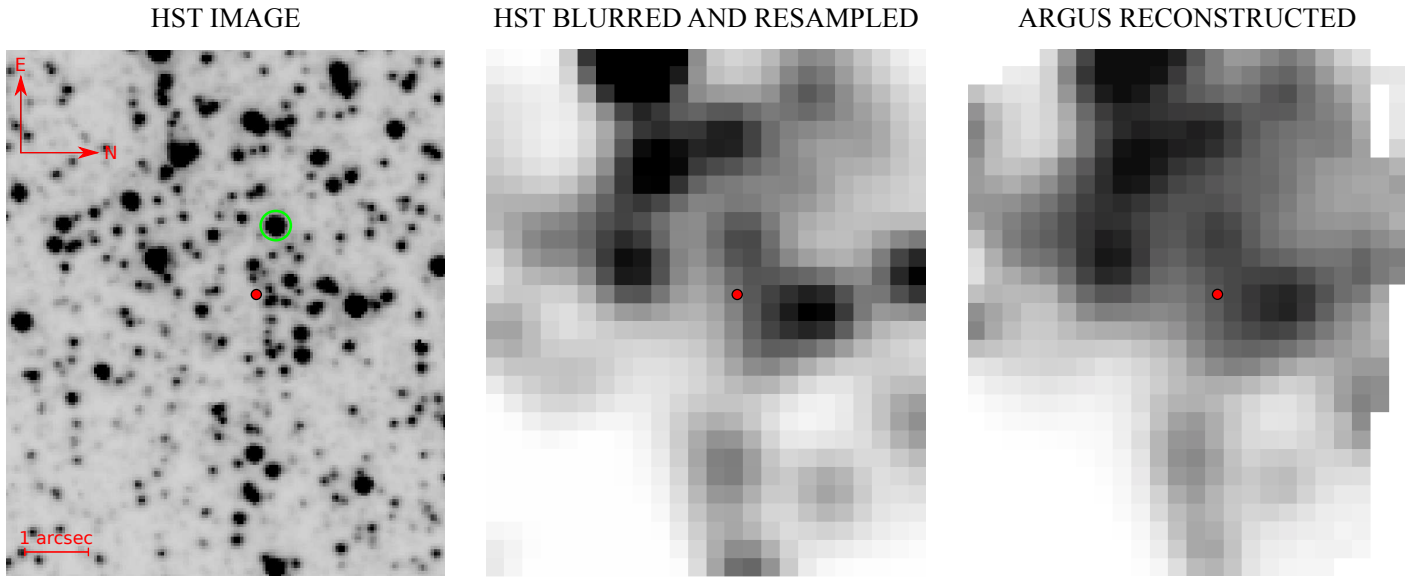}
      \caption{The ARGUS field of view: the HST image is shown on the left. The red circle marks the center, the green one the template star which was used. The same image but convolved with a Gaussian and resampled to the 0.3” pixel scale of the ARGUS array [center]. Compared to the actual reconstructed ARGUS pointing [right], it is clear that they are pointing to the same region.}
         \label{argus}
   \end{figure*}

         \begin{figure*}
   \centering
   \includegraphics[width=\textwidth]{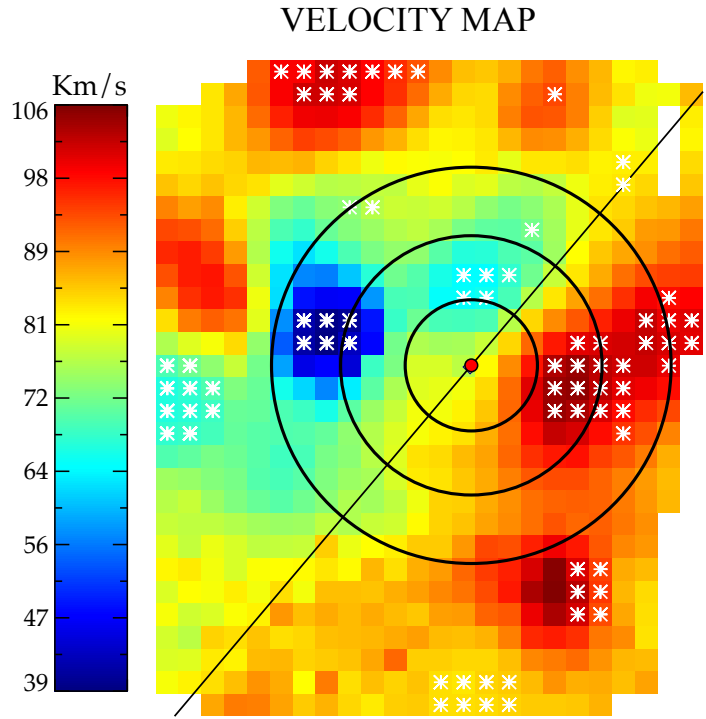}
      \caption{The velocity map of NGC 6388. As shown on the velocity scale, the blue spaxels indicate approaching velocities and the red ones receding. The white stars mark the spaxels which we excluded from the velocity dispersion measurement as they might suffer from shot noise. Also shown are the first three velocity dispersion bins to help visualize the binning method.}
         \label{vel}
   \end{figure*}

   
\section{Kinematics} \label{kin}

This section describes how we created a velocity map in order to check for rotation or other peculiar kinematic signals and how we derived a velocity-dispersion profile. This profile is then used to fit analytic models, described in the next section.


\subsection{Velocity map}

The reconstructed images of the three ARGUS pointings were matched to the HST image. With this information, the pointings were stitched together and each spectrum correlated to one position in the field of view. The resulting catalog of spectra and their coordinates allowed us to combine spectra in different bins. The combined pointing contains 24 $\times$ 29 spaxels. 

For each spaxel, the penalized pixel-fitting (pPXF) program developed by \cite{cappellari_2004} was used to derive the kinematics in that region. Figure \ref{argus} shows a) the field of view of the three pointings on the HST image, b) the field convolved with a Gaussian and resampled to the resolution of the ARGUS array and c) the reconstructed and combined ARGUS image. The corresponding velocity map is shown in Figure \ref{vel}. For every ARGUS pointing, a velocity map was derived before combining. The C pointing is a bit more noisy due to the fact that it only had one exposure. To check for systematic wavelength offsets we compared the derived velocities of the different pointings and exposures at overlapping spaxels. We found offsets of up to 1 $\mathrm{km\, s.^{-1}\, }$ We checked for systematic wavelength offsets between the pointings by cross-correlating their averaged sky spectra. The rms of the shifts measures 0.006 pixel (0.0018 \AA) which corresponds to a velocity shift of $\sim$ 0.1 $\mathrm{km\, s.^{-1}\, }$ We conclude that the velocity offsets are not due to uncertainties in the wavelength calibration and do not cause problems for our further analysis. Instead, the offsets are explained as follows:

i) Different pointings were dithered about half an ARGUS spaxel. This means that in every pointing a slightly different combination of stars contributes to the different spaxel, which could cause a different velocity measurement. ii) Errors in the reconstruction of the pointing positions could also cause velocity offsets between two positions. If the spaxels do not point at the exact same position we measure, similar to point i), different velocities. iii) Not all pointings had the same exposure time. This causes different signal-to-noise ratios for different pointings. This could also explain some variations in the velocity measurements with the pPXF routine. Note that for the velocity dispersion measurements, we grouped the individual spectra in large annuli. For that reason, the position offsets do not cause a problem for the further analysis as the  ``error" in position is not propagated into the velocity dispersion measurement as a systematic shift. 

We used a star from the central pointing (pointing A) as a velocity template. This has the advantage that it went through the same instrumental set-up as well as the same reduction steps as all other spectra and therefore the same sampling and wavelength calibration. The green circle in the HST image in Figure \ref{argus} (left panel) and the blue star in the color magnitude diagram (Figure \ref{cmd}) marks the star which was used. We also identified the brightest stars form the pointing in the CMD to make sure that the template and other dominating stars are not foreground stars (see Figure \ref{cmd}). In order to derive an absolute velocity scale, the line shifts of the templates were measured by fitting a Gaussian to each line and deriving the centroid. This was compared with the values of the Calcium Triplet in a rest frame and the average shift was calculated. We then corrected the radial velocity for the heliocentric reference frame.

\subsection{Shot noise corrections \label{snoise}} 

An obvious feature of the velocity map in Figure \ref{vel} is the bright blue spot in the middle left of the pointing which seems to dominate the blue (approaching) part of the map. We investigated whether we see a real rotation around the center or just one or a few bright stars with a peculiar velocity dominating their environment and with their light contaminating neighboring spaxels. Therefore, we tested the influence of each star on the adjacent spaxels. This allowed us to test if the derived velocity dispersion is representative of the entire population at a given radius or whether it is biased by a low number of stars, i.e. dominated by shot noise.

\begin{table}        
\centering
\caption{The four moments of the velocity distribution of NGC 6388.} 
\label{tab_moments}      
\begin{tabular}{c c c c c}
\hline \hline
\noalign{\smallskip}
  
$\log r$ 	& V & $\rm V_{\rm{RMS}}$& $h3$& $h4$ \\ 

[arcsec] 			& $[\mbox{km} / \mbox{s}]$ 	&$[\mbox{km} / \mbox{s}]$ 	&   & \\ 
\noalign{\smallskip}
\hline
\noalign{\smallskip}

$-0.05$ & $-1.6 \pm 2.9$ & $23.3 \pm 3.1$ & $-0.05 \pm 0.16$ & $-0.16 \pm 0.13$ \\
 $0.26$ & $-2.2 \pm 0.6$ & $25.7 \pm 1.8$ & $-0.07 \pm 0.01$ & $-0.10 \pm 0.01$ \\
 $0.43$ & $-0.1 \pm 0.6$ & $24.0 \pm 1.3$ & $-0.08 \pm 0.02$ & $-0.07 \pm 0.02$ \\
 $0.56$ & $ 2.9 \pm 0.6$ & $22.2 \pm 1.1$ & $-0.09 \pm 0.02$ & $-0.09 \pm 0.02$ \\
 $0.65$ & $ 6.4 \pm 0.6$ & $20.5 \pm 0.9$ & $-0.08 \pm 0.02$ & $-0.07 \pm 0.02$ \\
 $0.78$ & $ 4.6 \pm 0.8$ & $18.6 \pm 0.8$ & $-0.03 \pm 0.04$ & $-0.01 \pm 0.03$ \\
 $1.04$ & $ 3.5 \pm 0.8$ & $16.8 \pm 2.9$ & $-0.12 \pm 0.03$ & $-0.14 \pm 0.02$ \\
 $1.26$ & $ 6.7 \pm 1.0$ & $13.6 \pm 2.9$ & $-0.07 \pm 0.04$ & $-0.03 \pm 0.04$ \\

\noalign{\smallskip}
\hline
\end{tabular}
\end{table}

To perform this test, we considered our photometric catalog (described in section \ref{phot}) for the field of view covered by the ARGUS pointings. At every position of a star in the catalog, a two dimensional Gaussian was modeled with a standard deviation set to the seeing of the ground based observations ($\mathrm{FWHM}=0.9''$) and scaled to the total flux of the star. The next step was to measure the absolute amount and fraction of light that each star contributes to the surrounding spaxels. After computing these values for every star in the pointing, we had the following information for each spaxel: a) how many stars contribute to the light of that spaxel, and b) which fraction of the total light is contributed by each star, i.e. we determined whether the spectrum in a given spaxel was dominated by one or a few stars. The test showed that most of the spaxels contained meaningful contributions by more than 10 stars. Some spaxels, however, were dominated by a single star contributing more than 50 \% to the spaxel's light. For this reason, the contribution in percent of the brightest star was also derived by the program. We found out that the blue area in the left side of the velocity map of Figure \ref{vel} is not due to a single star. In fact this area in the velocity map corresponds to a group of at least 10 stars moving with 10 - 40 \kms with respect to the cluster systemic velocity.

\subsection{Velocity-dispersion profile}

To derive a radial velocity-dispersion profile for the stellar population of NGC 6388 it is necessary to bin the spectra accordingly. We divided the pointing into six independent angular bins, each of them with the same width of three ARGUS spaxels $(0.9'', \, 0.04 \, \mathrm{pc})$. To check the effect of the distribution of the bins on the final result, we tried different combinations of bins and bin distances as well as overlapping bins. We found no change in the global shape of the profile when using different bins. Therefore the choice of the bins was not critical, but in order to make an accurate error estimation, independent bins are more useful.

In each bin, all spectra of all exposures where combined with a sigma clipping algorithm to remove any remaining cosmic rays. Velocity and velocity-dispersion profiles were computed using the pPXF method applied to the binned spectra. The velocity map in Figure \ref{vel} shows the dynamics in the cluster center. A rotation-like or a shearing signature is visible, but it seems to be a very local phenomenon (within $3'', \, 0.15 \ \mathrm{pc}$). In earlier attempts, we split the pointing in two halves (tilted line in figure \ref{vel}) in order check for consistency and symmetry on both sides. The velocity dispersion was then derived separately for the binned spectra on each side. Both sides show a rise in the velocity-dispersion profile but differ in their shape and absolute values from each other. However, treating both sides separately would not properly take into account a possible rotation at the center and therefore not exactly represent the observed data. We decided to use the total radial profile over the 360 degree bins and measure the second moment $\rm V_{\rm{RMS}}=\sqrt{\rm V_{\rm{rot}}^2+\sigma^2}$, with the rotation velocity $\rm V_{\rm{rot}}$ and the velocity dispersion $\sigma$. The reason why we chose to analyze V$_{\rm{RMS}}$ instead of $\sigma$ is twofold. First, the Jeans models require the V$_{\rm{RMS}}$ rather than the pure velocity dispersion as an input. Second, the broadening of the line we measured represents V$_{\rm{RMS}}$ and not $\sigma$. The velocity dispersion can be obtained by subtracting the rotation velocity from this quantity. Determining the rotation is difficult due to the large shot noise introduced by the small number of spatial elements in the central region. The second moment, however, is robustly measured since we average over all angles. For simplicity we refer to the V$_{\rm{RMS}}$ profile as the velocity-dispersion profile in our study.

In addition to the central pointings, we derived kinematics for regions further out using the small IFU measurements, which were scattered at larger radii around the cluster. Unfortunately, the surface brightness of the cluster drops quickly with radius resulting in a low signal-to-noise ratio for the IFUs most distant from the center. These could therefore not be used for further analysis. Only the two innermost positions showed reasonable signal, so that these two pointings could be used as two separated data points at $11''$ and $18''$ radius, respectively. The disadvantage of the small IFU fields (20 spaxels for a total field-of-view of $3'' \times 2''$) is that these values are very affected by shot noise since only a few stars fall into the small field-of-view. Consequently, these points show larger errors than the rest of the profile. 

Further, we estimated the radial velocity of the cluster in the heliocentric reference frame. We combined all spectra in the pointing and measured the velocity relative to the velocity of the template. This value was then corrected for the motion of the template and the heliocentric velocity and results in a value of  $\rm V_r = (80.6 \pm 0.5)$ \kms which is in good agreement with the value from \cite{harris_1996} $\rm V_r = (81.2 \pm 1.2)$ \kms.

We ran Monte Carlo simulations to estimate the error on V$_{\rm{RMS}}$ due to shot noise. From the routine described in section \ref{snoise}, we knew exactly how many stars are contributing (\textit{nStars}) and how many spaxels are summed (\textit{nSpax}) in each bin. We took all stars detected in each bin and their corresponding magnitude. Each of the stars was then assigned a velocity chosen from a Gaussian velocity distribution with a fixed dispersion. We used our template spectrum and created \textit{nSpax} spaxel by averaging \textit{nStars} flux weighted and velocity shifted spectra. The resulting spaxels were then normalized, combined and the kinematics measured with pPXF (as for the original data). After 1000 realizations for each bin, we obtained the shot noise errors from the spread of the measured velocity dispersions. For the outer IFU pointings, we extrapolated the surface density to larger radii and performed the same Monte Carlo simulations as for the inner pointing,  with random magnitudes drawn from the magnitude distribution in this region. The errors for the velocity and the higher moments were derived by applying Monte Carlo simulations to the spectrum itself. This was done by repeating the measurement for 100 different realizations of data by adding noise to the original spectra. \citep[see ][section 3.4]{cappellari_2004}

\begin{figure*}
   \centering
 \includegraphics[width=\textwidth]{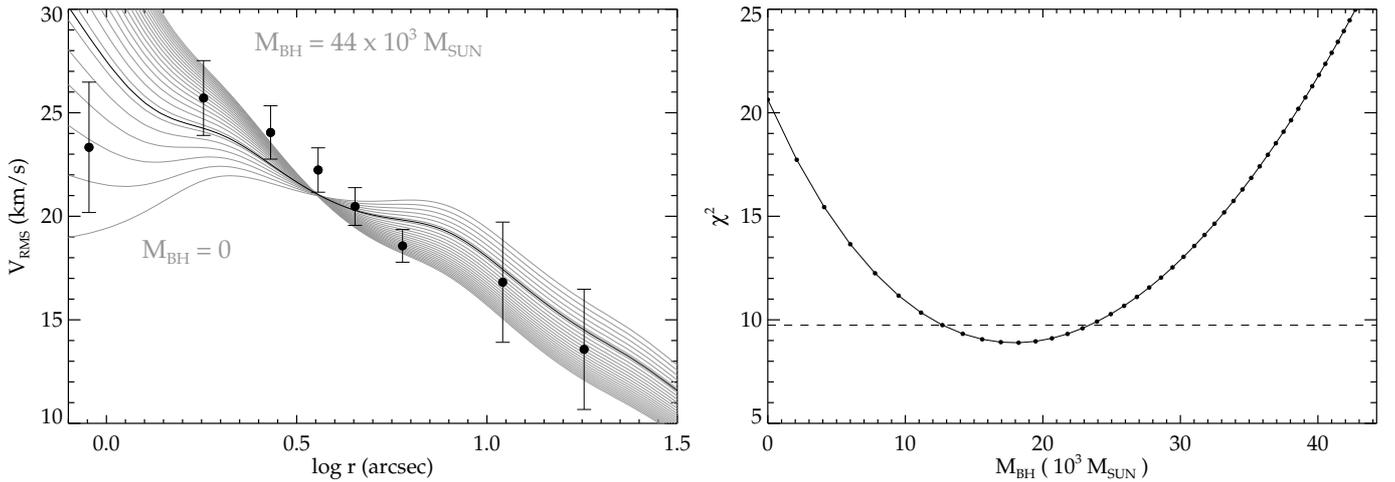}
      \caption{Different isotropic ($\beta = 0$ fixed) Jeans models applied to the kinematic data from our pointing. The M/L is fitted to the data by scaling the profile to the data points. The final values are obtained by finding the minimum of $\chi^2$ while varying the black-hole mass: $\beta= 0.0, \, M_{BH}=18.24 \times 10^3 M_{\odot}, \, M/L_V = 1.65, \, \chi^2_{min}=8.89$. The plot on the left panel shows the models together with the data and the best fit. Labeled are also the black holes masses of the two enclosing models. The solid black line marks the best fit. The right panel shows the $\chi^2$ values of every model as well as the $\Delta \chi^2 = 1$ line (dashed line). }
         \label{models}
   \end{figure*}

The resulting profile is displayed in Figure \ref{models}. The plot shows the second moment of the velocity distribution V$_{\rm{RMS}}$. Except for the innermost point, a clear rise of the profile towards the center is visible. The highest point reaches more than 25 \kms before it drops quickly below 20 \kms at larger radii. In table \ref{tab_moments} we list the results of the kinematic measurements. The first column lists the radii of the bins. The following columns show the central velocities of each bin in the reference frame of the cluster, the second moment V$_{\rm{RMS}}$ as well as higher velocity moments h3 and h4. It is conspicuous that all h4 moments tend to have negative values. This hints at a lack of radial orbits in the central region of the globular cluster.  

For general purposes we also determined the effective velocity dispersion $\sigma_e$ as described in \cite{gultekin_2009}. This is defined by:

 \begin{equation} \sigma_e^2 = \frac{\int_{0}^{R_e} \rm V_{\rm{RMS}}^2 \, I(r) \, dr}{\int_{0}^{R_e} I(r) \, dr} \end{equation} \label{sig_e}
 
Where $R_e$ stands for the half light radius ($\sim 40''$) and $I(r)$ the surface brightness profile. We extrapolated our kinematic data to the half light radius since our furthest out data point only reaches a radius of $18 ''$. This results in an effective velocity dispersion of $\sigma_e= (18.9 \pm 0.3)$ \kms which is in perfect agreement with the value derived by \cite{pryor_1993} listed in table \ref{tab_6388}.

\section{Dynamical models \label{jeans}} 

The main goal of this work is to compare the derived kinematics and light profile with a set of simple analytical models in order to test whether NGC 6388 is likely to host an intermediate-mass black hole in its center. We used Jeans models as implemented and described in \cite{cappellari_2008}.

\subsection{Isotropic and anisotropic models}

The first input for the Jeans models is the surface brightness profile in order to estimate the 3-D density distribution in the cluster. Given the fact that the density can only be observed in a projected way, the profiles have to be deprojected. One way of doing this is by applying the multi-Gaussian expansion (MGE) method developed by \cite{emsellem_1994}. The basic approach of this method is to parametrize the projected surface brightness with a sum of Gaussians since the deprojection of a Gaussian function results again in a Gaussian.

To apply this parametrization and to compare results of the Jeans equation to our data, we used the Jeans Anisotropic MGE (JAM) dynamical models for stellar kinematics of spherical and axisymmetric galaxies, as well as the multi-Gaussian expansion implementation developed by \cite{cappellari_2002,cappellari_2008}. The IDL routines provided by M. Cappellari\footnote{Available at http://purl.org/cappellari/idl} enable the modeling of the surface brightness profile and fitting the observed velocity data and the mass-to-light ratio at the same time. We used our surface brightness profile in combination with a spherical model with different values of anisotropy and constant $M/L_V$ setups along the radius of the cluster.

In Figure \ref{models}, we plot the isotropic model data comparison for our velocity-dispersion profile. The JAM program does not actually fit the model to the data but rather calculates the shape of the second moment curve from the surface brightness profile, deconvolves the profile with the PSF of the IFU observations, and then scales it to an average value of the kinematic data in order to derive the $M/L_V$. This explains why all trial models meet in one point. The thick black line marks the model with the lowest $\chi^2$ value and therefore the best fit.

\begin{figure}
   \centering
   \includegraphics[width=0.5\textwidth]{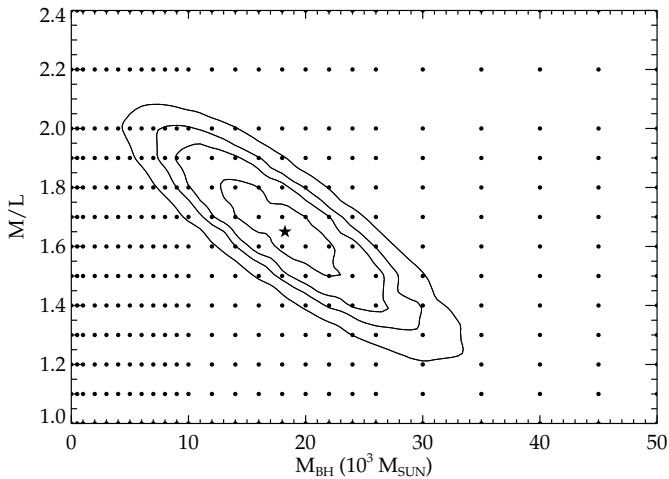}
      \caption{The contours of the $\chi^2$ as a function of black-hole mass and mass-to-light ratio. Each point represents a particular isotropic model. The plotted contours are $\Delta \chi^2 = 1.0, 2.7, 4.0$ and $6.6$ implying a confidence of $68, 90, 95$ and $99$ percent.}
         \label{chi_cont}
   \end{figure}

With the given surface brightness profile and without a central black hole, the models predict a drop in velocity dispersion towards the central region. As a final result, we used the $\chi^2$ statistics of the fit to estimate an error. This results in: $M_{\bullet} = (18 \pm 6) \times 10^3 M_{\odot}$ and $M/L_V = (1.7 \pm 0.2) \ M_{\odot}/L_{\odot}$. Figure \ref{chi_cont} shows the contour plot of $\chi^2$ as a function of black-hole mass and mass-to-light ratio over a grid of isotropic models (black points). The contours represent $\Delta \chi^2= 1.0, 2.7, 4.0$ and $6.6$ which correspond to a confidence of $68, 90, 95$ and $99$ percent for 1 degree of freedom (marginalized). This implies that for an isotropic model and this specific surface brightness profile the models predict a black hole of at least $M_{\bullet} = 5 \times 10^3 M_{\odot}$ with a confidence of 99 percent.

We also tested whether anisotropic Jeans models would result in a significantly better fit. To do this, we repeated the model fitting with $\beta \neq 0$ values. We found a decrease of $\chi^2$ for a rising $\beta$ down to $\chi^2=4.21$ for $\beta=0.5$ with a lower black-hole mass of $M_{\bullet}= 5.3 \times 10^3 M_{\odot}$. However, this requires a very high anisotropy of $\beta = 1 - \overline{\rm V_{\theta}^2} / \overline{\rm V_{r}^2}=0.5$ which is expected to be unstable in the center of a relaxed globular cluster such as NGC 6388. The anisotropy will be discussed in more detail in section \ref{sec_aniso}.

\subsection{Error estimation}

In section \ref{phot_sb} we introduced the different surface brightness profiles for NGC 6388. The \cite{trager_1995} profile contains the most data points and extends to large radii ($\sim 4'$). However, the inner regions (important for our dynamical purposes) are not as well sampled by the ground based observations used by Trager et al. The profile by NG06 was derived by measuring integrated light on a WFPC2 image using a bi-weight estimator and covers the inner regions of the cluster very well. To calibrate the profiles to the correct absolute values NG06 also adjusted it to the Trager et al. profile. The third profile was obtained by \cite{lanzoni_2007} by computing the average of the counts per pixel in each bin of their ACS/HRC images. Since it was not clear in the paper how they calibrated their photometry, this profile was also shifted (by a small amount of $0.09 \, \mathrm{mag}$) to overlap with the Trager et al. profile. As a test, we used each of these profiles as an input for the Jeans models and found similar black-hole masses varying from $M_{\bullet} = (25 \pm 6) \times 10^3 M_{\odot}$ for the \cite{trager_1995} profile to $M_{\bullet} = (11 \pm 5) \times 10^3 M_{\odot}$ for the \cite{lanzoni_2007} data. The result from the profile provided by NG06 is identical to the fit of our profile in black-hole mass and slightly lower in the mass-to-light ratio ($M/L_V = (1.5 \pm 0.2) \ M_{\odot}/L_{\odot}$). Again we find the lowest $\chi^2$ value (and therefore the best fit) for an anisotropic model of $\beta=0.5$. In this case the observations can be explained without any black hole for all profiles. As mentioned in the previous section this rather unlikely configuration will be discussed in more detail in the next subsection.

\begin{figure}
   \centering
   \includegraphics[width=0.5 \textwidth]{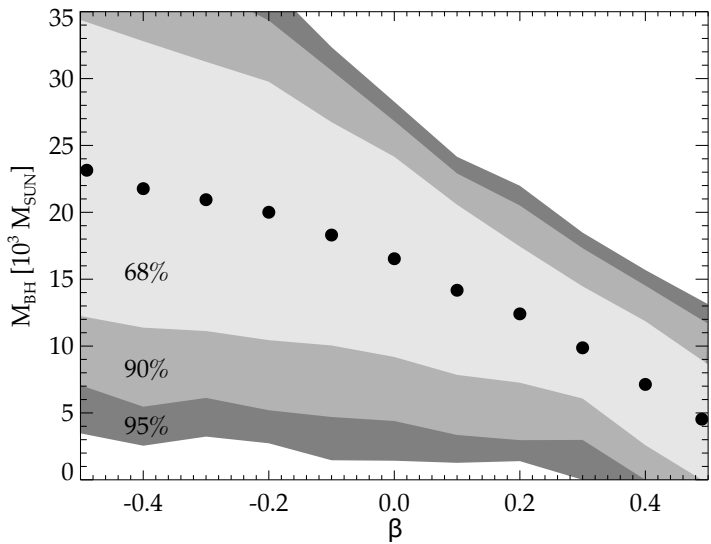}
      \caption{The result of the Monte Carlo simulations of the surface brightness profile. For each fixed $\beta$ value 1000 realizations of the light profile have been performed. The black dots mark the average value of the best fit black-hole mass. The shaded contours represent the $68, 90$ and $95$ percent confidence limits.}
         \label{montecarlo}
   \end{figure}

From our studies of the surface brightness profiles, we see that they yield similar, but not identical results. The shape of the SB profile predicts the shape of the velocity-dispersion profile. Therefore, it is crucial to test the effect of variation of the surface brightness profile in the inner regions. To perform this, we run Monte Carlo simulations on the six innermost points of the profile. We used 1000 runs and a range of $\beta$ between $-0.5$ and $0.5$. The results are displayed in figure \ref{montecarlo}. The black points mark the mean value of the derived black-hole masses. The shaded contours represent the $68, 90$ and $95$ percent confidence limits. The mean black-hole mass decreases at higher $\beta$ values since a radial anisotropy can mimic the effect of a central black hole. Nevertheless, for an isotropic case, we have a black hole detection within a mass range of  $ 17_{-7}^{+8} \times 10^3 M_{\odot}$ and a global mass-to-light ratio of $M/L_V = 1.6_{-0.2}^{+0.2} \ M_{\odot}/L_{\odot}$ (errors are the 68 percent confidence limits). We derived the total error (resulting from kinematic and photometric data) by again, running Monte Carlo simulations and varying both, the velocity-dispersion profile and the surface brightness profile at the same time. As a final result for the uncertainties of our method we derived $\delta M_{\bullet} = 9 \times 10^3  M_{\odot}, \, \delta M/L_V = 0.3 \ M_{\odot}/L_{\odot}  $ . This shows, that a significant percentage of the error ($ \sim 75 \%$) in the context of isotropic Jeans models results from the uncertainties of the surface brightness profile.

\subsection{Anisotropy in relaxed globular clusters \label{sec_aniso}}

In the previous sections we have shown, that the strong significance for a black hole with isotropic models vanishes when using anisotropic models. The anisotropic Jeans models show a better fit for a model with much lower black-hole masses or without a black hole requiring an anisotropy of $\beta=0.5$ or higher. Thus, it is necessary to investigate the probability of such a kinematic configuration in our cluster. We run N-body simulations for $30000$ equal-mass stars, distributed according to a Plummer model. The initial anisotropy was created by a cold collapse, which resulted in a cluster that was slightly anisotropic in its core and had increasing anisotropy further out rising to $\beta=0.9$ outside the half-mass radius. Furthermore, we took the tidal field of the galaxy into account by letting the cluster move on a circular orbit around a point-mass galaxy. The ratio between tidal radius and core radius was set to 1:10 as it is the case for most globular clusters as well as for NGC 6388. We then let the cluster evolve over a time scale of several relaxations times and calculated the anisotropy for snapshots of the model. 

\begin{figure}
   \centering
   \includegraphics[width=0.5 \textwidth]{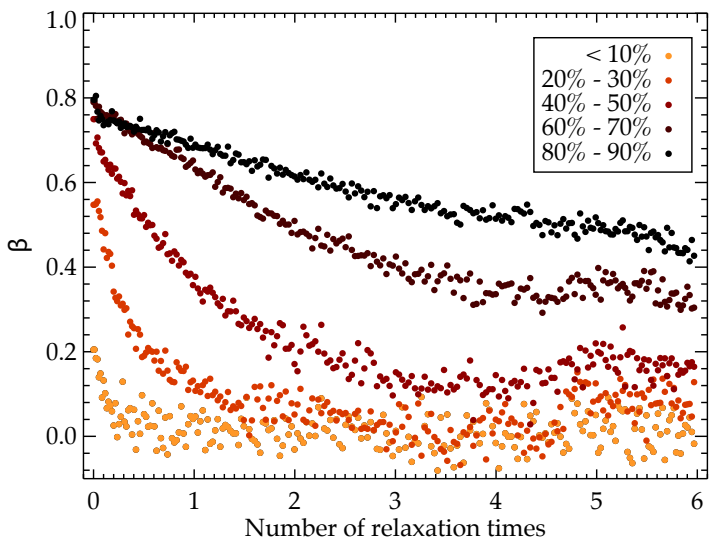}
      \caption{The anisotropy $\beta$ as a function of time for different regions in the cluster. The different colors represent different percentages of the cluster total light/mass going from the inner part of less than $10 \%$ to the outskirts of the cluster with $90 \%$ light/mass enclosed. }
         \label{aniso}
   \end{figure}

Figure \ref{aniso} shows the result of these tests. The different colors represent different areas in the cluster. Important for our case are the kinematics inside the half-light radius (curves with $< 50 \%$ light/mass enclosed in figure \ref{aniso}). The system relaxes very quickly in the central regions. After six relaxation times the inner regions of the cluster are almost isotropic with $\beta$ values below $0.2$. The relaxation time of NGC 6388 is of order $10^9 yrs$ \citep[at the half-mass radius, ][]{harris_1996} which implies that the cluster is about $10$ relaxation times old. This shows, that high anisotropies (such as the the ones needed from our best fit models) in NGC 6388 are not stable and would have vanished over a short time scale. Therefore, we limit our discussion to the results from our isotropic models.

\subsection{Remaining concerns \label{concerns}}

In our analysis we used the assumption of a constant $M/L_V$ for the entire cluster. In reality, this ratio can vary with radius and it may not be well described with a single value if the cluster is mass segregated. The rise in the velocity-dispersion profile could then be mimicked by a dark remnant cluster at the center. Such a scenario is expected in core collapsed clusters which display surface brightness profiles with logarithmic slopes of -0.8 or higher \citep{baumgardt_2005}. 
We looked at the deprojected light-density  and mass-density profiles resulting from the surface brightness profile and the velocity-dispersion profile. The mass-density profile from the kinematics drops with $\varpropto r^{-2}$ which would be expected from a core collapsed cluster. But the light profile is shallow and does not support that hypotheses: with a concentration of $c=1.7$ and a logarithmic slope of the surface brightness profile of $\alpha = -0.28 \pm 0.08$ NGC 6388 does not show any signs of core collapse in the distribution of the visible stars. We compared these profiles with simulated core collapsed or mass segregated clusters \citep[described in][]{baumgardt_2003} and were not able to find a good agreement with either shape of the light profile or slope of the mass density of the cluster. We currently consider the presence of a dark remnant cluster unlikely but we will perform detailed n-body simulations with a variable M/L in the near future.


\section{Summary and conclusions}\label{con}


We derived the mass of a potential intermediate-mass black hole at the center of the globular cluster NGC 6388 by analyzing spectroscopic and photometric data. With a set of HST images, the photometric center of the cluster was redetermined and the result of NG06 confirmed. Furthermore, a color magnitude diagram as well as a surface brightness profile, built from a combination of star counts and integrated light, were produced. The spectra from the ground-based integral-field unit ARGUS were reduced and analyzed in order to create a velocity map and a velocity-dispersion profile. In the velocity map, we found signatures of rotation or at least complex dynamics in the inner three arcseconds (0.15 pc) of the cluster. We derived a velocity-dispersion profile by summing all spectra into radial bins and applying a penalized pixel fitting method. 

Using the surface brightness profile as an input for spherical Jeans equations, a model velocity-dispersion profile was obtained. We ran several isotropic models with different black-hole masses and scaled them to the observed data in order to measure the mass-to-light ratio. Using $\chi^2$ statistics, we were able to find the model which fits the observed data best. We run Monte Carlo simulations on the inner points of the surface brightness profile in order to estimate the errors resulting from the particular choice of a light profile. From this, we determined the black-hole mass and the mass-to-light ratio as well as the scatter of these two results. The final results, with $68 \%$ confidence limits are: A black-hole mass of $ (17 \pm 9) \times 10^3 M_{\odot}$ and a global mass-to-light ratio of $M/L_V = (1.6 \pm 0.3) \ M_{\odot}/L_{\odot}$. In addition, we run anisotropic models. The confidence of black hole detection is decreasing rapidly with increasing radial anisotropy. However, using N-body simulations, we found that in a relaxed cluster such as NGC 6388 an anisotropy higher than $\beta=0.2$ inside the half mass radius will be erased within a few relaxation times.

\begin{figure}
   \centering
   \includegraphics[width=0.5\textwidth]{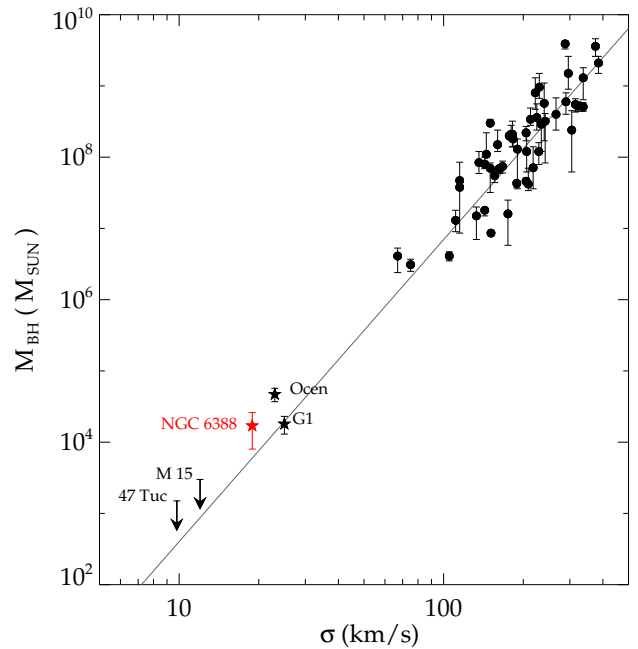}
      \caption{The $M_{\bullet} - \sigma$ relation for galaxies at the high mass range (filled circles) and the first globular clusters with potential black hole detection (filled stars). The slope of the line $\log (M_{\bullet}/M_{\odot}) = \alpha + \beta \, \log ( \sigma / 200 \mathrm{km\, s^{-1}\, })$ with $(\alpha, \beta) = (8.12 \pm 0.08, 4.24 \pm 0.41)$ was taken from \cite{gultekin_2009}. The mass of the black holes of $\omega$ Centauri, G1 and the upper limits of M15 and 47 Tuc were obtained by \cite{noyola_2010}, \cite{gebhardt_2005}, \cite{bosch_2006} and \cite{47tuc}, respectively. Overplotted is the result of NGC 6388 with our derived effective velocity dispersion $\sigma_e = 18.9$ \kms.}
         \label{m_sigma}
   \end{figure}

With respect to the black-hole mass estimate of $5.7 \times 10^3 \ M_{\odot}$ by \cite{lanzoni_2007}, which they derived from photometry alone, our derived black-hole mass is larger by a factor of three with our surface brightness profile profile. Using their light profile, we obtained a black-hole mass of $M_{\bullet} = (11 \pm 5) \times 10^3 M_{\odot}$ which includes their value within one sigma error. \cite{cseh_2010} obtained deep radio observations of the inner regions of NGC 6388 and discussed  different accretion scenarios for a possible black hole. They found an upper limit for the black-hole mass of $\sim 1500 M_{\odot}$ since no radio source was detected at the location of any Chandra X-ray sources. However, inserting our black-hole mass in equation 5 of \cite{cseh_2010} and using their assumptions as well as the Chandra X-ray luminosity of the central region, results in minimal but possible accretion rates and conversion efficiency ($\varepsilon \eta \sim 10^{-7}$). Considering the fact that supermassive black holes have masses not much higher than $0.2$ \% of the mass of their host systems \citep{marconi_2003,haering_2004}, our mass with $0.9 \%$ seems to be higher than the predictions for larger systems. Globular clusters in contrast lose much of their mass during their evolution. This could naturally result in higher values of black-hole mass - host system mass ratios. Due to the complicated dynamics, the $1 \sigma$ uncertainties are 40 \% for our black-hole mass and 10 \% for the derived $M/L_V$. Figure \ref{m_sigma} shows the position of NGC 6388 in the black-hole mass velocity dispersion relation. With our derived effective velocity dispersion of $\sigma_e = (18.9 \pm 0.3)$ \kms the results for NGC 6388 seem to coincide with the prediction made by the $M_{\bullet} - \sigma$ relation. 

The mass-to-light ratio of $M/L_V = (1.7 \pm 0.3) \, M_{\odot}/L_{\odot}$ derived in this work is consistent within the error bars with the dynamical results of \citet[$M/L_V\sim 1.8 $]{McLaughlin_2005} but slightly lower than the value predicted by stellar population models. According also to \cite{McLaughlin_2005}, we would expect a value of $M/L_V = (2.55 \pm 0.28) \, M_{\odot}/L_{\odot}$ for $\mathrm{[Fe/H]}=-0.6 \,  \mathrm{dex}$ and  $(13 \pm 2) \, \mathrm{Gyr}$. \citet{baumgardt_2003} have shown that dynamical evolution of star clusters causes a depletion of low-mass stars from the cluster. Since these contribute little light, the $M/L_V$ value drops as the cluster evolves. Dynamical evolution could therefore explain part of the discrepancy between theoretical and observed $M/L_V$ values. 

Summarizing, it can be said that the globular cluster NGC 6388 shows a variety of interesting features in its photometric properties (e.g. the extended horizontal branch) as well as in its kinematic properties (e.g. the high central velocity dispersion or the rotation like signature in the velocity map). In this work we investigated the possibility of the existence of an intermediate-mass black hole at the center. With simple analytical models we were already able to reproduce the shape of the observed velocity-dispersion profile very well if the cluster hosts an IMBH. Future work would be to compute detailed N-body simulations of NGC 6388 in order to verify whether a dark cluster of remnants at the center would be an alternative. Additionally, it is crucial to obtain kinematic data at larger radii to constrain the mass-to-light ratio and the models further out. Also proper motions for the central regions would further constrain the black hole hypothesis.

The study of black holes in globular clusters is currently limited to a handful of studies. For this reason, it is necessary to probe a large sample of globular clusters for central massive objects in order to move the field of IMBHs from ``whether" to ``under which circumstances do" globular clusters host them. Thus tying this field to the one of nuclear clusters and super-massive black holes. From the experience we gathered with NGC 6388, we can say that the dynamics of globular clusters is not as simple as commonly assumed and by searching for intermediate-mass black holes, we will also be able to get a deeper insight into the dynamics of globular clusters in general.

\begin{acknowledgements}
This research was supported by the DFG cluster of excellence Origin and Structure of the Universe (www.universe-cluster.de). We also thank Nadine Neumayer for constructive feedback and inspiring discussions, Annalisa Calamida for her support in the photometric analysis and Giuseppina Battaglia for helping with the sky subtraction. H.B. acknowledges support from the Australian Research Council through Future Fellowship grant FT0991052. We thank the anonymous referee for constructive comments and encouraging further analysis concerning anisotropy and error estimation.
\end{acknowledgements}

\bibliographystyle{aa}
\bibliography{ref}

\end{document}